# Momentum Transfer to Nanoobjects between Parallel Heated Plates

S. Hardt[a,1], S. Tiwari[2] and A. Klar[2,3]

[1]*Institut für Nano- und Mikroprozesstechnik, Leibniz Universität Hannover, D-30167 Hannover, Germany*

[2]*Fraunhofer ITWM, D-67663Kaiserslautern, Germany*

[3]*Fachbereich Mathematik, TU Kaiserslautern, D-67663 Kaiserslautern, Germany*

A small-scale, trapezoidal rigid body in the gas-filled gap between two parallel plates at different temperatures is considered. An analytical expression for the force onto the body in the direction parallel to the plates valid for an infinite Knudsen number is derived. Simultaneously, Monte-Carlo simulations are performed allowing to extend the analysis to Knudsen numbers of the order of one. The numerical and the analytical results show excellent agreement, indicating that a temperature gradient orthogonal to the plates can induce a significant force in parallel direction. This force is only slightly reduced when a Knudsen number of one is considered. In the future, the discovered effect may be exploited for the conversion of thermal into mechanical energy in nanomachines.

[a] Corresponding author. Email: hardt@nmp.uni-hannover.de, Tel.: +49-511-7622278, Fax: +49-511-7622167

On the scale below one micrometer internal gas flows at standard conditions exhibit rarefaction effects which are due to the fact that collisions of molecules with the boundary of the flow domain start to play an important role [1]. In such small-scale systems transport phenomena that are unknown from or unimportant in macroscopic devices can appear [2,3]. Corresponding effects need to be taken into account when designing nanomachines and -devices, and may be exploited to create novel operation and actuation principles. One example is thermal creep, being the induction of a mass flow by a temperature gradient [4]. Another example is the thermophoretic force on a rigid body immersed in a gas of nonuniform temperature [5]. By contrast to these effects where a momentum transfer is induced parallel or antiparallel to the temperature gradient, in this article we wish to show how the motion of a small solid body *perpendicular to* a temperature gradient can be induced. The situation we consider is an arrangement of parallel plates between which a rigid body of trapezoidal cross-section can move, as depicted in Fig. 1(a). The gap is filled with a gas and it is assumed that the surfaces of the plates are kept at constant, but generally different temperatures. The goal of our studies is to compute the force on the rigid body (the "nanoobject") due to collisions with surrounding gas molecules. For a gas at zero Knudsen number (Kn), defined as the ratio of the molecular mean free path and the width of the gap, it can easily be shown that the force components parallel to the walls of the plates due to collisions with the right and the left wall of the rigid body exactly cancel, such that the net force on the body is zero. The question whether or not this situation changes when the Knudsen number is larger than 1 is the main objective of this study.

For the force calculations it is assumed that infinitely many identical nanoobjects separated by a distance $L$ (measured at the bottom plate) are arranged periodically within the gap. It is then sufficient to consider the geometrical domain between two successive bodies, as depicted in Fig 1(b). The case of a single body in an infinitely long gap is recovered by taking the limit $L \to \infty$. The geometry of the domain is fully defined by the parameters $L$, $d$,

$\alpha_r$ and $\alpha_l$. For later convenience the walls of the trapezoidal domain are labeled with numbers 0, …, 3, where 0 corresponds to the left and 2 to the right wall of the nanoobject. All analytical calculations reported in this letter have been performed in two spatial dimensions.

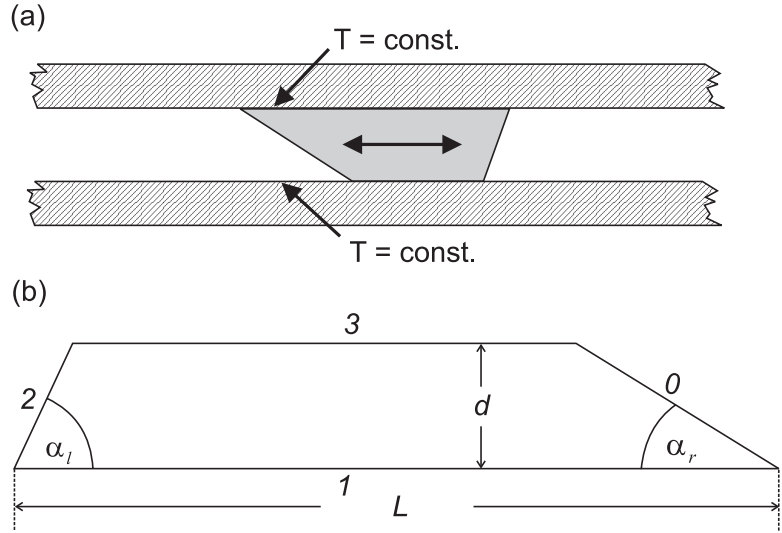

FIG. 1. Schematic of a nanoobject of trapezoidal shape moving between parallel plates kept at constant temperature (a) and sketch of the geometrical domain for which the force calculations have been performed (b).

As a first step, we investigate the stationary-state configuration (in the sense of a statistical average) of an ensemble of gas molecules in the limit of free molecular flow. The molecular velocity distribution is then solely determined by the boundary conditions at the walls and the geometry of the domain. Before going back to the specific situation shown in Fig. 1 let us more generally consider a convex, simply connected two-dimensional domain $\Omega$ as depicted in Fig. 2. Inside the domain with boundary $d\Omega$ gas molecules can move freely. We introduce an arc-length parameter $s \in [0, s_{max}]$ that measures the length of the boundary $d\Omega$. The number of molecules leaving an infinitesimal segment of the boundary $ds_1$ and impacting on another segment $ds_2$ is given by $\dot{n}_\vartheta(s_1, \vartheta) ds_1 d\vartheta$, where $\vartheta$ is the angle between the vector connecting $s_1$ and $s_2$ and the unit normal on $ds_1$. Furthermore, we introduce the total number of molecules per arc-length emitted from a wall segment as

$$\dot{n}(s) = \int_{-\pi/2}^{\pi/2} \dot{n}_g(s_1,\vartheta)d\vartheta \qquad (1)$$

Because of mass conservation for a stationary-state molecular distribution in the domain $\Omega$, we require that the total number of molecules per unit time being emitted from each wall segment $ds$ equals the sum of the molecules per unit time being emitted from all other wall segment and impacting on $ds$. Mathematically, this is expressed as

$$\dot{n}(s) = \int_0^{s_{max}} \chi(s',s)\dot{n}(s')ds', \qquad (2)$$

where $\chi(s',s)ds$ is the probability that a molecule emitted from $s'$ impacts on $ds$. $\chi$ usually displays a complicated functional dependence on the geometry of the domain $\Omega$ and the angular dependence $\dot{n}_g$ of the emission. Eq. (2) is a Fredholm integral equation of the second kind and can be solved numerically using an approximation for the integral and an iteration scheme.

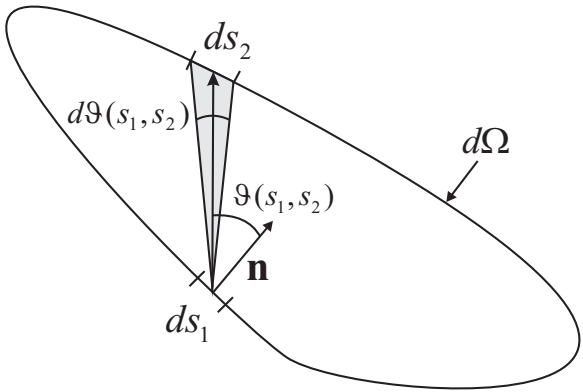

FIG. 2. Domain $\Omega$ with boundary $d\Omega$ in which gas molecules are moving.

In the following we simplify the problem by assuming that complete accommodation of molecules reflected from the walls of the domain takes place, which means that a molecule impacting on a wall loses the "memory" of its velocity before the collision and is reflected diffusively with a Maxwell distribution reflecting the local temperature at the wall. The corresponding boundary condition will be denoted as diffuse-reflection boundary condition

(DRBC). Specifically, the angular distribution of the diffusively reflected molecules is given by

$$\dot{n}_g(s, \vartheta) = \frac{1}{2}\dot{n}(s)\cos\vartheta. \qquad (3)$$

It should be pointed out that the DRBC is clearly not valid in general, especially at higher energies [6]. Experimental results, however, indicate that thermal and momentum accommodation coefficients of more than 0.95 can be reached in interactions of molecules with surfaces [7]. Thus, the DRBC should at least come close to a number of realistic situations. In order to capture a broader class of gas-surface interaction phenomena, some more complicated empirical models have been developed [8,9].

The DRBC greatly simplifies the solution of Eq. (2), as will be shown in the following. The principle of detailed balance expresses that in a steady-state situation a process and its reverse occur with the same probability. Translated to our problem it follows that

$$\dot{n}(s')\chi(s',s) = \dot{n}(s)\chi(s,s'). \qquad (4)$$

In addition, we exploit an analogy between free molecular gas dynamics and the problem of radiation heat transfer between surfaces. Within that analogy the quantity $\chi(s',s)ds$ takes the role of a view factor for radiation going from $ds'$ to $ds$. If and only if the angular dependence of that radiation is given by Eq. (3), the reciprocity relation

$$\chi(s',s) = \chi(s,s') \qquad (5)$$

is valid [10]. Thus, we arrive at the important identity $\dot{n}(s') = \dot{n}(s)$, concluding that the number of molecules emitted from $d\Omega$ per unit length and unit time is constant over the whole boundary.

We will now use these results to compute the forces on the walls of the nanoobject, areas 0 and 2 in Fig. 1(b), in the limit $L \to \infty$. In that limit the contribution of the interaction between walls 0 and 2 vanishes and only the interaction with molecules emitted from the horizontal walls has to be taken into account. Consider, without loss of generality, the momentum transfer on a segment $ds$ of wall 0, as depicted in Fig. 3. By applying the principle of detailed balance we find that the number of molecules impacting on $ds$ from a direction **r** within an angle interval $[\alpha, \alpha + d\alpha]$ is given by

$$\frac{1}{2}\dot{n}(s)\cos\vartheta = \frac{1}{2}\dot{n}(s)\sin\alpha. \qquad (6)$$

Taking into account that molecules impacting under an angle $\alpha < \alpha_r$ ($\alpha > \alpha_r$) must originate from wall 3 (from wall 1), the force in $x$-direction on segment $ds$ is given as

$$f_x^{(0)}(s)ds = \frac{\dot{n}(s)ds}{2}\left(\int_0^{\alpha_r}\overline{p}_x^{(3)}\sin\alpha\,d\alpha + \int_{\alpha_r}^{\pi}\overline{p}_x^{(1)}\sin\alpha\,d\alpha\right), \qquad (7)$$

where $\overline{p}_x^{(1)}, \overline{p}_x^{(3)}$ are the $x$-components of the momenta of molecules impacting under an angle $\alpha$ from walls 1, 3. The bar over the momenta indicates that an average over a statistical distribution has to be computed. With $\overline{p}_x^{(i)} = \overline{p}^{(i)}\cos(\alpha_r - \alpha)$, where $\overline{p}^{(i)}$ is the mean absolute value of the momentum, and making use of $\dot{n}$ being a constant, it follows in a straightforward manner that the total $x$-momentum per unit time of molecules impacting on wall 0 is given by

$$F_x^{(0)} = \frac{1}{4}\dot{n}d\left((\pi - \alpha_r)\overline{p}^{(1)} + \alpha_r\overline{p}^{(3)}\right). \qquad (8)$$

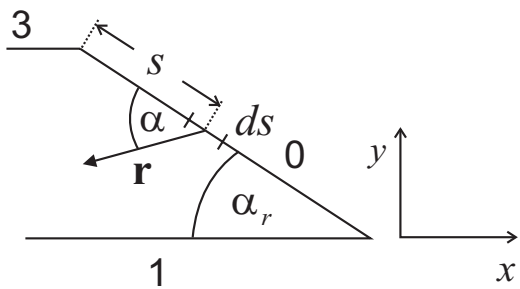

FIG. 3. Segment $ds$ on wall 0 with direction vector parametrized by $\alpha$.

In order to compute the total force on the nanoobject, the contribution of molecules impacting on wall 2 has to be subtracted from this expression. In addition, also the recoil momenta of molecules leaving the two walls have to be taken into account. From their Maxwellian distribution together with the angular dependence of Eq. (3) it is easy to see that the recoil contributions from walls 0 and 2 to the mean force exactly cancel if the temperature in the nanoobject is only a function of $y$ (cf. Fig. 3). This assumption should be fulfilled with good accuracy for a body between parallel plates at fixed temperature through which heat is conducted. Thus, the total force in $x$-direction on the nanoobject is given by

$$F_x = \frac{1}{4}\dot{n}d(\alpha_l - \alpha_r)(\bar{p}^{(1)} - \bar{p}^{(3)}) \qquad (9)$$

The mean momenta appearing in this expression can be computed from the wall temperatures via the Maxwellian. Clearly the momenta are only different if the walls are at different temperatures. This shows that work can only be generated between reservoirs of different temperature, in accordance with the second law of thermodynamics. In addition, the force on the nanoobject depends on its geometry through the angles $\alpha_r$ and $\alpha_l$ and vanishes if the angles are equal. We can conclude that while at vanishing Knudsen number no force is exerted, there is a considerable force in the free molecular regime that depends on the wall temperatures and on the geometry of the body.

A natural scale to compare the force of Eq. (9) with is given by the force per unit length exerted on walls 1 and 3. This is simply the pressure force from the enclosed gas molecules which is easy to compute in the limit $L \to \infty$. Dividing $F_x/d$ from Eq. (9) by this quantity yields a dimensionless pressure given by

$$P_x = \frac{\alpha_l - \alpha_r}{\pi} \frac{\bar{p}^{(1)} - \bar{p}^{(3)}}{\bar{p}^{(1)} + \bar{p}^{(3)}}. \qquad (10)$$

In order to check the analytical results and to extend the analysis to Kn ≈ 1, a Monte-Carlo method (a variant of the DSMC method [11] developed in [12, 13]) was applied to solve the Boltzmann equation in the computational domain depicted in Fig. 1. The method is based on the time splitting of the Boltzmann equation. Introducing fractional steps one solves first the free transport equation (the collisionless Boltzmann equation) for one time step. During the free flow boundary and interface conditions are taken into account. In a second step (the collision step) the homogeneous Boltzmann equation without the transport term is solved. To simulate this equation by a particle method an explicit Euler step is performed. The result is then used in the next time step as the new initial condition for the free flow. To solve the homogeneous Boltzmann equation the key point is to find an efficient particle approximation of the product of distribution functions in the Boltzmann collision operator given only an approximation of the distribution function itself. To guarantee positivity of the distribution function during the collision step a restriction on the time step proportional to the mean free path is needed. That means that the method becomes exceedingly expensive for small Knudsen numbers, but is fast for the cases considered here. For more details about the solution procedure we refer to the above cited references.

In the present calculation the number of time steps is $2 \cdot 10^6$. Sampling of the results is done over the last $8 \cdot 10^5$ time steps. The grid cells are quadratic and of size $0.1d$. In the case Kn = ∞ the same grid as for Kn = 1 is chosen. Initially the particles are generated according to the Maxwellian distribution with a temperature of 300 K and a vanishing mean velocity. For each test case between 40 and 80 particles per cell have been initialized.

It should be noted that, in contrast to the analytical approach presented above, the Monte-Carlo method is based on a 3D description. In relation to the wall reflection boundary condition of Eq. (3) this means that the angle $\vartheta$ takes the role of the azimuthal angle, while the reflected particles are equally distributed over the polar angle $\phi$.

The first set of numerical computations was performed with the DRBC, $\alpha_l = 90°$, wall temperatures $T_1 = 300$ K, $T_2 = 450$ K, a temperature profile on walls 0 and 2 linearly interpolating between the temperatures on walls 1 and 3, and a number of particles $N_p$ of about 80,000. The angle $\alpha_r$ was varied between 30° and 90°. Fig. 4 shows the negative dimensionless pressure as a function of $\Delta\alpha = \alpha_l - \alpha_r$ for two different Knudsen numbers compared with the analytical result valid for Kn = ∞. The error bars were computed from an extrapolation towards $N_p = \infty$ based on three different sets of computations with particles numbers of about 40,000, 60,000 and 80,000.

The numerical data points on the left side of the figure obtained for Kn = ∞ show excellent agreement with the analytical results. Apparently the restriction to two spatial dimensions in the analytical model does not introduce any notable deviations from the three-dimensional case. Furthermore, the pressure on the nanoobject is somewhat reduced when a Knudsen number of 1 is considered. This is expected, since in this case only particles emitted within a distance of about $d$ can reach the walls of the nanoobject. However, even at Kn = 1 the analytical result still remains a reasonable approximation to the force on the nanoobject. The fact that the dimensionless pressure vanishes at $\Delta\alpha = 0$ shows that the effect observed is related to breaking the reflection symmetry with respect to the $x$-axis.

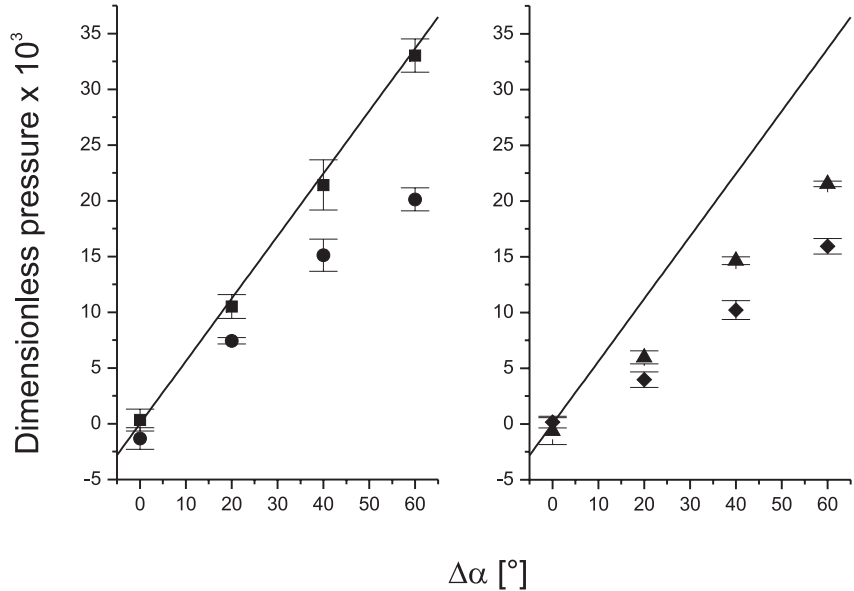

FIG. 4. Negative dimensionless pressure as a function of $\Delta\alpha = \alpha_l - \alpha_r$. Line: analytical result (DRBC); Squares: Kn = ∞ (DRBC); Circles: Kn = 1 (DRBC); Triangles: Kn = ∞ (mixed BC); Diamonds: Kn = 1 (mixed BC). The symbols represent numerical results.

As stated above, the DRBC can be expected to be valid approximately in a significant number of cases, but is clearly not the most general relationship of practical importance. In order to check how the wall boundary conditions influence the results, the DRBC was mixed with a specular reflection boundary condition in such a way that both types of reflections occur with an equal probability at all walls. In a specular reflection the momentum component of a particle tangential to the wall is conserved, whereas the normal component is inverted [7].

The results displayed in Fig. 4 show that the negative dimensionless pressure is significantly reduced if the mixed BC is used instead of the DRBC. About 60-70 % of the original values are reached. Naively, one would expect a pressure reduction to 50 %, since in the case of purely specular reflection at the walls and initialization of a particle ensemble with $T_{ini} = (T_1 + T_3)/2$ the Monte-Carlo simulation yields a vanishing force onto the nanoobject. However, there are subtle interaction effects between the two types of wall reflections which result in an augmentation of the exerted force. A qualitative explanation for this phenomenon can be given as follows. For the DRBC the contributions of the recoil momenta from particles reflected from the walls of the nanoobject (#0 and #2) exactly cancel. If, however, two subsequent wall collisions of different type come into play, this is no longer true. In this context it can either happen that a particle interacting specularly with the walls of the parallel plates is reflected diffusively from the walls of the nanoobject in the next step or vice versa. In the former case it is easy to show that the net recoil momentum onto the nanoobject exactly vanishes. In the latter case this no longer holds, as can be seen from considering a situation where all the reflections from walls 1 and 3 are diffuse, while all the reflections from walls 0 and 2 are specular. In this case the force onto the nanoobject originating from the recoil momenta can be derived in a straightforward manner as

$$F_x^{(recoil)} = \frac{1}{2}\dot{n}d\left(\frac{\alpha_l - \alpha_r}{2} + \sin\alpha_r \cos\alpha_r - \sin\alpha_l \cos\alpha_l\right)\left(\bar{p}^{(1)} - \bar{p}^{(3)}\right) \quad (11)$$

For the geometry parameters considered here this force points into the same direction as the force of Eq. (9), thus showing that the interaction of diffuse and specular wall reflections yields an augmentation of the force obtained for the mixed BC beyond the naively expected value.

In summary, we have considered the problem of a trapezoidal body arranged between parallel plates at different temperature. From an analytical solution for the distribution of gas molecules between the plates in the limit Kn = ∞ an expression for the force onto the body depending on the wall temperatures and the geometry was derived. Re-expressing the result as a dimensionless pressure shows that heat transfer between the plates results in a significant momentum transfer onto the nanoobject. The analytical curve was corroborated by Monte-Carlo simulations, yielding excellent agreement. The effect discovered can be interpreted as a novel mechanism for conversion of thermal into mechanical energy and may find applications in future nanosystems.